\documentclass[twocolumn,prl,floatfix,superscriptaddress,showpacs]{revtex4}

\usepackage{epsfig}

\begin{document}

\title{Influence of Strain on the Kinetics of Phase Transitions in Solids}
\author{Efim A. Brener}
\affiliation{Institut f\"ur Festk\"orperforschung, Forschungszentrum J\"ulich, D-52425 J\"ulich, Germany}
\author{V.I. Marchenko}
\affiliation{P.L. Kapitza Institute for Physical Problems, RAS 119334, Kosygina 2, Moscow, Russia}
\author{R. Spatschek}
\affiliation{Institut f\"ur Festk\"orperforschung, Forschungszentrum J\"ulich, D-52425 J\"ulich, Germany}
\date{\today}
\begin{abstract}
We consider a sharp interface kinetic model of phase transitions accompanied by elastic strain, together with its phase-field realization.
Quantitative results for the steady-state growth of a new phase in a strip geometry are obtained and different pattern formation processes in this system are investigated.
\end{abstract}

\maketitle

Many magnetic, superconducting and structural phase transitions in solids are accompanied by small lattice distortions which lead to the presence of elastic deformations.
For each phase transition a characteristic deformation can be assigned, i.e., the deformation acquired by the new phase relative to the initial phase in the absence of external forces.
In some cases, these effects are of minor influence and can be ignored, but nevertheless for many applications the elastic strain causes qualitatively new and observable effects.
The influence on the thermodynamics of transitions between different phases has been thoroughly discussed in the literature (for a review see \cite{R,Kh}, and, for more recent developments, e.g.~\cite{delaey01} and references therein).

One of the well known consequences is a thermodynamic elastic hysteresis, i.e.~the splitting of the phase equilibrium point into two points, the points of the direct and inverse transition.
It is mainly due to the coherency at the interphase boundary, meaning that the lattice layers remains continuous through the boundary.
Correspondingly, the hysteresis disappears without interface coherency \cite{BM}.
Despite of the general understanding, some features of such transformations are still unclear, or at least under debate.
For example, a distinctive two-phase equilibrium is established in the system within a certain temperature interval.
The nature of this phenomenon is difficult to understand from the standpoint of ordinary thermodynamic equilibrium concepts:
during the process of martensitic transformations the composition of the phases does not change, and thus in such systems only one phase can be stable at a given temperature.

However, the systematic theoretical study of the growth {\it kinetics} of such phase transitions accompanied by a lattice strain is much less advanced.
In real systems the influence of elastic strain is often screened by many other effects, for example, by inhomogeneous compositions and temperature distributions, the Mullins-Sekerka instability, crystal anisotropy, polycrystalline structures, etc.
Here, a phase-field modeling of such complicated systems can lead to qualitative descriptions of the kinetics of phase transitions in solids (see, for example, \cite{steinbach06} and references therein).
The main purpose of this Letter is to develop a ``minimum'' kinetic model from which even quantitative results concerning the influence of strain effects can be obtained.

We start from the thermodynamical description of our model.
The free energy density of an initial phase is
\begin{equation}\label{F1}
F_1=F_1^0+\frac{\lambda}{2}u_{ii}^2+\mu u_{ik}^2,
\end{equation}
where $F_1^0$ is the free energy density without elastic effects, $u_{ik}$ is the strain tensor, $\lambda$ and $\mu$ are the elastic moduli of isotropic linear elasticity.
The free energy density of a new phase is
\begin{equation}\label{F2}
F_2=F_2^0+\frac{\lambda}{2}\left(u_{ii}-u^0_{ii}\right)^2+\mu\left(u_{ik}-u_{ik}^0\right)^2,
\end{equation}
with $u_{ik}^0$ being a characteristic lattice strain assigned to the phase transition.

Let us consider the simplest case, ${u^0_{ik}=\varepsilon\delta_{ik}}$, at first.
We assume that the elastic effects are small, ${\varepsilon\ll 1}$, and neglect the difference between the elastic coefficients in the two phases.
Since in our description the reference state for both phases is the undeformed initial phase (see Eqs.~(\ref{F1}) and (\ref{F2})), the coherency condition reads ${\bf u}^{(1)}={\bf u}^{(2)}$, where ${\bf u}$ is the displacement vector.
The superscripts $(1)$ and $(2)$ refer to the initial and the newly created phase respectively.
Mechanical equilibrium at the interface demands ${\sigma^{(1)}_{nn}=\sigma^{(2)}_{nn}}$ and ${\sigma^{(1)}_{n\tau}=\sigma^{(2)}_{n\tau}}$.
Here, the stress tensor is given by $\sigma_{ik}=\frac{1}{2}(\partial F/\partial u_{ik}+\partial F/\partial u_{ki})$;
the indices $n$ and $\tau$ denote the normal and tangential directions with respect to the interface.
The condition of phase equilibrium requires the continuity of a new potential $\tilde F$ across the flat interface \cite{P},
\begin{equation}\label{PA}
{\tilde F}=F-\sigma_{nn}u_{nn}-2\sigma_{n\tau}u_{n\tau}.
\end{equation}
In the general case of curved interfaces also the surface energy $\gamma$ should be taken into account, and the phase equilibrium condition reads in the case of isotropic surface energy
\begin{equation}\label{PK}
{\tilde F}_1-{\tilde F}_2-\gamma {\cal K}=0,
\end{equation}
where ${\cal K}$ is the curvature of the interface.

A critical nucleus of the new phase inside an unbounded initial phase can exist only if
\begin{equation}\label{critic}
F^0_1-F^0_2>\frac{E\varepsilon^2}{1-\nu},
\end{equation}
where $E$ and $\nu$ are Young's modulus and Poisson ratio respectively.
This condition corresponds to the elastic hysteresis mentioned above.
It can be obtained using the analogy of this elastic problem to the problem of thermal expansion for a given temperature field, as described in \cite{LL}.

\begin{figure}
\begin{center}
\epsfig{file=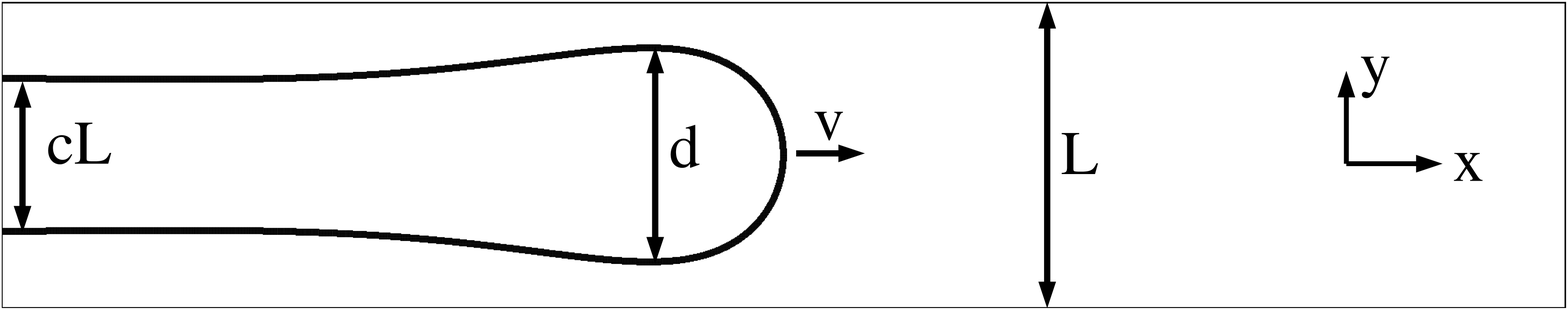, width=7.5cm}
\epsfig{file=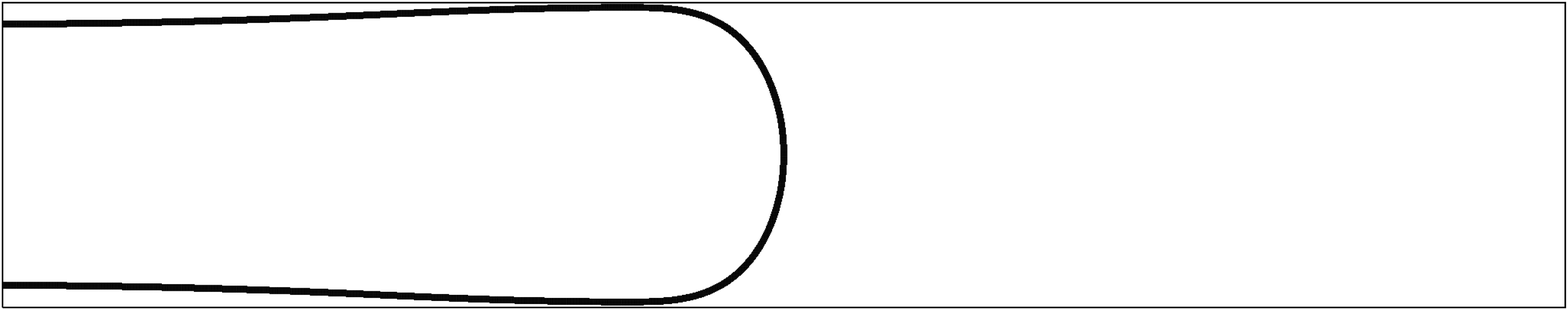, width=7.5cm}
\epsfig{file=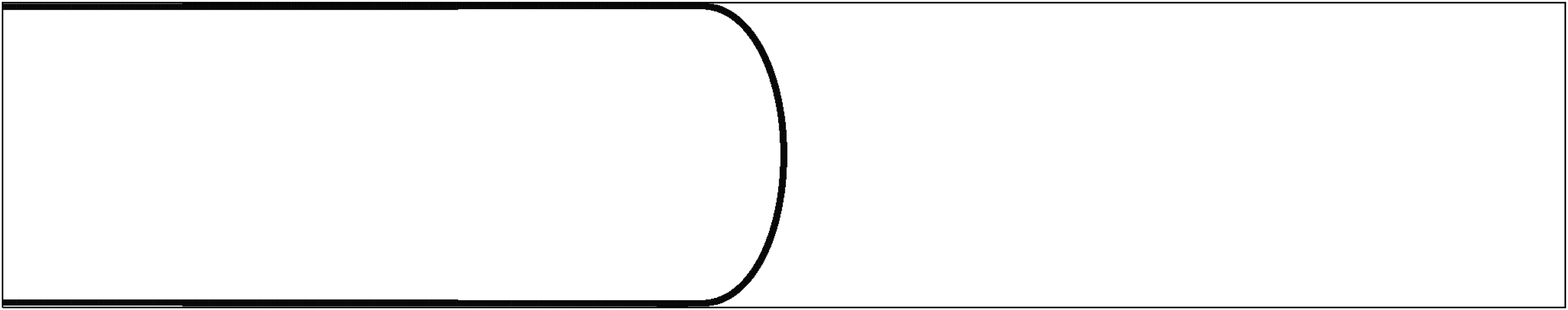, width=7.5cm}
\caption{Shapes of propagating fingers calculated for three values of the driving force. Top: $\Delta=0.5<\Delta_t$. Center: $\Delta=0.85>\Delta_t$. Bottom: $\Delta=1.05$. The Poisson ratio is $\nu=1/3$ and $L/L^*=10$.}
\label{fig1}
\end{center}
\end{figure}

We discuss a simple strip configuration which allows the steady-state growth of an elastic ``finger'' consisting of the new phase (see Fig.~\ref{fig1}).
The unstrained elastic strip of width $L$ is attached to fixed grips at the upper and lower boundary (${\bf u}=0$ there) and initially composed of the reference phase.
We discuss a two-dimensional elastic problem using plane strain conditions ($u_z=0$).
Also, we assume the complete wetting of the walls by the initial phase.
Thus, the new phase avoids a direct touching of the walls (see Fig.~\ref{fig1}).
Far ahead of the propagating finger the initial phase remains unstrained.
In contrast, far behind the tip a phase coexistence is possible within a certain parameter interval near the transition temperature, which is due to elastic effects.
In this region the only nonvanishing component of the displacement vector is $u_y$.
The strain tensors are constant in both phases and their nonzero components are connected to each other by the relation ${(1-c)u^{(1)}_{yy}+cu^{(2)}_{yy}=0}$, in order to fulfill the conditions ${u_y=0}$ on the walls and ${u^{(1)}_y=u^{(2)}_y}$ at the interface.
Here, $c$ is the volume fraction of the new phase.
Then, the mechanical equilibrium condition gives
\begin{equation}\label{strain}
u^{(1)}_{yy}=-\frac{1+\nu}{1-\nu}c\varepsilon,\qquad u^{(2)}_{yy}=\frac{1+\nu}{1-\nu}(1-c)\varepsilon.
\end{equation}
Taking into account the phase equilibrium condition Eq.~(\ref{PK}) we find the volume fraction of the new phase,
\begin{equation}\label{c}
c=\Delta=\frac{1-2\nu}{1+\nu}\left[\frac{1-\nu}{E\varepsilon^2}(F^0_1-F^0_2)-1\right],
\end{equation}
which defines a dimensionless driving force $\Delta$ for this process.
Then the parameter range for coexistence is $0<\Delta<1$.
The total energy gain of this two-phase configuration compared to the unstrained initial phase is
\begin{equation}\label{gain}
\Delta{\cal F}=L[F^0_1-cF_2-(1-c)F_1]-2\gamma.
\end{equation}
Finally, using Eq.~(\ref{c}) we find
\begin{equation}\label{gain1}
\Delta{\cal F}=2\gamma\left(\frac{\Delta^2}{\Delta_L^2}-1\right),
\end{equation}
where
\begin{equation}\label{cL}
\Delta^2_L=\frac{L^*}{L},\qquad L^*=4\frac{(1-2\nu)(1-\nu)\gamma}{(1+\nu)E\varepsilon^2}.
\end{equation}
The finger grows if ${\Delta{\cal F}>0},$ or, equivalently, if ${\Delta>\Delta_L}$.

Two remarks are in order.
First, we obtain the two-phase structure only because we take into account the elastic effects and use the fixed volume boundary condition.
For stress free boundaries, we would obtain $c=1$ for any driving force $F_1^0-F_2^0$ above the threshold (\ref{critic}) and $c=0$ below the threshold. Second, the value of the driving force for the transition can be controlled not only by temperature, but also by external strain.
In particular, if the strip of width $L$ is stretched by $\delta L$, the homogeneous term $\delta L/L$ has to be added to the strain $u_{yy}$.
This eventually leads only to a renormalization of the driving force,
\begin{equation}\label{ren}
F_1^0-F_2^0 \rightarrow F_1^0-F_2^0+ \frac{E\varepsilon}{1-2\nu}\frac{\delta L}{L}.
\end{equation}

Following our general aim to develop a minimum kinetic model, we assume that the growth is controlled only by interface kinetics.
Then, the local equation of motion of the interface reads
\begin{equation}\label{v}
v_n=\kappa({\tilde F}_1-{\tilde F}_2-\gamma {\cal K}),
\end{equation}
where $v_n$ is the normal velocity and $\kappa$ a kinetic coefficient.

The conservation of energy requires that the excess $\Delta{\cal F}$ is compensated by dissipation at the interface.
This leads to a relation between the growth velocity and the driving force for the process,
\begin{equation}\label{velocity}
\frac{v}{v_0L}\int n_x^2ds=\Delta^2-\Delta_L^2,
\end{equation}
where $v$ is the steady-state velocity of the finger and
\begin{equation}\label{v0}
v_0=2\kappa\gamma/L^*=\frac{\kappa (1+\nu)E\varepsilon^2}{2(1-2\nu)(1-\nu)}
\end{equation}
is the characteristic velocity scale for this system;
$n_x$ is the projection of the interface normal on the growth direction $x$ and the integration is performed along the interface.
The dimensionless quantity $\int n_x^2ds/L$ is a complicated function of the parameters $\Delta$, $\Delta_L$, $\nu$;
in the case of dynamical elasticity it also depends on the ratio $v_0/c_s$ where $c_s$ is the shear wave speed.
Near the equilibrium point $\Delta=\Delta_L$, the growth velocity behaves as
$v/v_0\propto \Delta-\Delta_L$.
These results are valid for the dimensionless driving force $\Delta<1$.
For $\Delta>1$ the fraction of the second phase becomes unity and Eq.~(\ref{velocity}) should be replaced by
\begin{equation}\label{velosity1}
\frac{v}{v_0L}\int n_x^2ds=2\Delta -1-\Delta_L^2.
\end{equation}

In order to obtain quantitative results for this problem we use a phase field code together with elastodynamics to describe phase transformations accompanied by stress, which we developed recently \cite{spat06}.
Let $\phi$ denote the phase field with values $\phi=1$ for the initial phase and $\phi=0$ for the new phase.
The energy density contribution is $F=F_1h(\phi)+F_2[1-h(\phi)]$,
where the switching function $h(\phi)=\phi^2(3-2\phi)$ interpolates between the phases.
The gradient energy is $F_s(\phi)=3\gamma \xi (\nabla\phi)^2/2$ with the interface width $\xi$.
Finally, $F_{dw}=6\gamma\phi^2(1-\phi)^2/\xi$ is the double well potential.
Thus the total energy functional is
${\cal F} = \int dV \left( F+F_s+F_{dw}\right)$.
The elastodynamic equations are derived from the energy by variation with respect to the displacements $u_i$,
$\rho \ddot{u}_i=-\delta {\cal F}/\delta u_i$,
where $\rho$ is the mass density.
The dissipative phase field dynamics follows from
$\dot{\phi} = -(\kappa/3\xi) \delta{\cal F}/\delta\phi$.
These equations lead in the limit ${\xi\to 0}$ to the  sharp interface description above.
For the case of static elasticity, this was proved in \cite{Kassner01}.

For the numerical realization, we employ explicit representations of both the elastodynamic equations and the phase field dynamics.
The elastic displacements are defined on a staggered grid \cite{Virieux86}.
We shift the grid horizontally in order to keep the propagating tip always in the center of the strip;
this allows to study steady-state growth in moderately large systems.
The intrinsic length scale $L^*$ is chosen to be larger than the phase-field interface width, ${L^*=8\xi}$, and ${\xi=5\,\Delta x}$, where ${\Delta x}$ is the numerical lattice unit.
We have performed calculations for the Poisson ratio $\nu=1/3$ and for two values of the dimensionless strip width $L/L^{*}=10$ and $L/L^{*}=20$, or equivalently, for two values of the parameter $\Delta_L\approx 0.32$ and $\Delta_L\approx 0.22$.
All simulations are conducted on the parallel computer JUBL operated at the Research Center J\"ulich.

First of all, we have checked that the asymptotic conditions in the two-phase region far behind the tip, Eqs.~(\ref{strain}) and (\ref{c}), are reproduced by our numerics with high precision.
Though the appearance of the Asaro-Tiller-Grinfeld instability \cite{AT,Grinfeld} might be naively expected in the tail region because of the presence of non-hydrostatic stresses, this effect is never observed in the simulations.
In the limit $L\to\infty$, we have checked analytically that the system is stable.
We have also confirmed the predicted renormalization of the driving force (\ref{ren}) due to an external strain $\delta L/L$.
Next, we have investigated the properties of growing elastic fingers for different driving forces, $F_1^0-F_2^0$.
After a transient regime the finger always reaches the steady-state configuration.
Characteristic shapes of the stationary fingers are presented in Fig.~\ref{fig1}.
The finger selects a symmetrical shape even if the initial configuration was vertically off-centered, i.e.~the symmetrical configuration is stable.
We note the existence of the ``bubble'' in the tip region (top panel of Fig.~\ref{fig1}).
The bubble touches the walls at a specific value ${\Delta=\Delta_t<1}$.
With further increase of the driving force the touching region smoothly increases and diverges at $\Delta=1$ (see middle and bottom panel of Fig.~\ref{fig1}).


\begin{figure}
\begin{center}
\epsfig{file=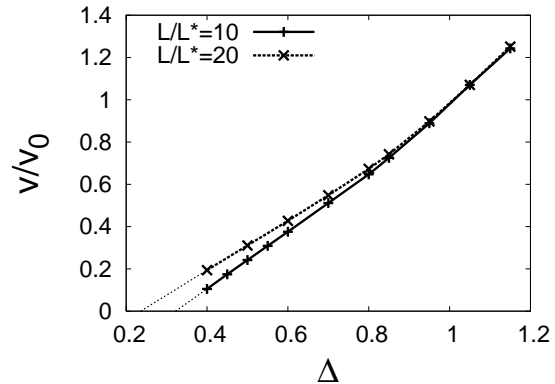, width=7.5cm}
\caption{The dimensionless growth velocity $v/v_0$ as a function of the dimensionless driving force $\Delta$.}
\label{fig3}
\end{center}
\end{figure}

The dependence of the dimensionless steady-state growth velocity $v/v_0$ on the dimensionless driving force $\Delta$ is presented in Fig.~\ref{fig3}.
The velocity rises from zero at ${\Delta=\Delta_L}$ with increasing driving force.
It turns out that the dependence is almost linear in a wide range, ${\Delta_L<\Delta<\Delta_t}$.
This is a slightly unexpected result, because the linear behavior is anticipated to be valid only in the close vicinity of the equilibrium point ${\Delta=\Delta_L}$.
At both critical points, ${\Delta=\Delta_t}$ and ${\Delta=1}$, the growth velocity is a continuous function of the driving force.

All the results presented so far have been obtained for the characteristic velocity scale $v_0$ being much smaller than shear wave speed, i.e.~in the limit of static elasticity.
We have performed additional runs with higher characteristic velocities such that $v/c_s\approx 1/2$, where dynamical effects are relevant.
For the same driving force, the dimensionless velocity $v/v_0$ decreases compared to the static elasticity limit, and the size of the bubble grows in order to reach the same dissipation according to Eq.~(\ref{velocity}).

Let us shortly discuss another simple example of transitions involving shear strain in hexagonal crystals.
For the transitions lowering the symmetry from $C_6$ to $C_2$ the shear strain in a basic plane appears.
For simplicity we neglect all other possible strains with higher (axial) symmetry.
We assume that the crystal is attached to two parallel walls as before.
Let the principal axis $C_6$ be oriented in $z$ direction.
By proper choice of the crystal orientation around the main axis in the initial phase, we obtain the new phase in three possible states having the following nonvanishing components of the strain tensor $u_{ik}^0$;
\begin{equation}\label{u6}
u^0_{xx}=-u^0_{yy}=\varepsilon\cos2\theta,\qquad u^0_{xy}=\varepsilon\sin2\theta,
\end{equation}
where the angle ${\theta=0,\pm 2\pi/3}$.
Because the elasticity of hexagonal crystals is axisymmetric in harmonic approximation and ${u_{iz}^0=u_{iz}=0}$ for the discussed problem, we can use Eqs.~(\ref{F1}) and (\ref{F2}) for the energy densities of the two phases (see e.g.~\cite{LL}).
The moduli of the effective isotropic elasticity, $\lambda$ and $\mu$, can be expressed in terms of the elastic constants of the original hexagonal crystal.

\begin{figure}
\begin{center}
\epsfig{file=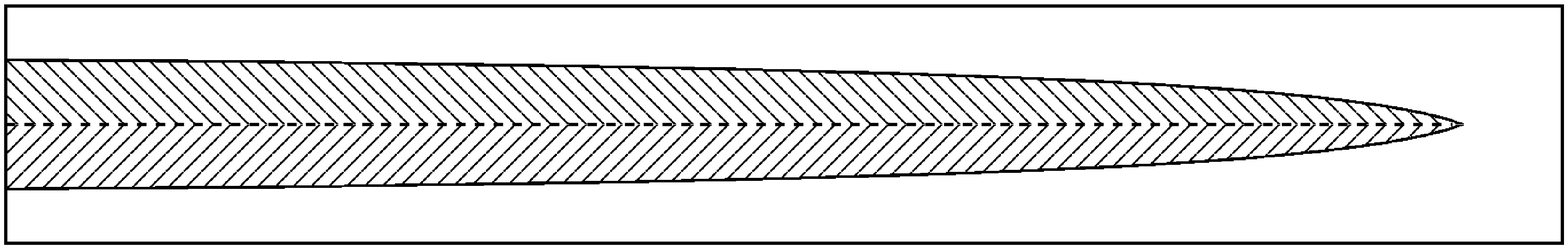, width=7.5cm}
\epsfig{file=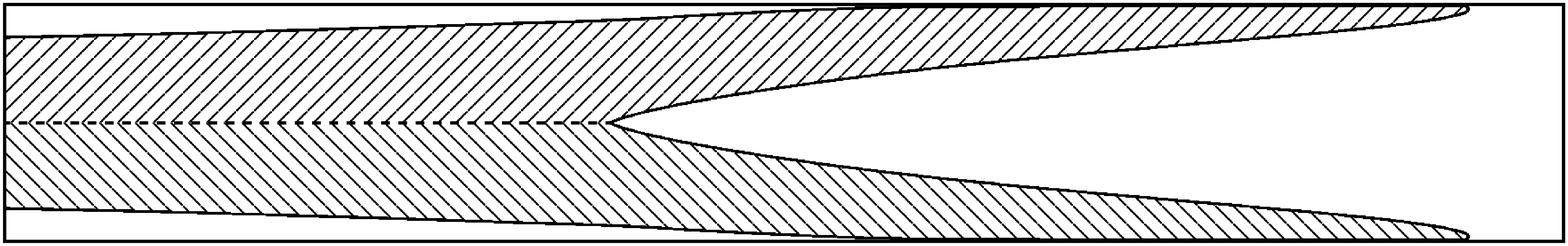, width=7.5cm}
\caption{Growth of bicrystal patterns for ${L/L^*=5}$. The shading illustrates the orientation of the shear strain $u_{xy}$. Forward slashes correspond to $\theta=-2\pi/3$, backslashes to $\theta=2\pi/3$. The strip lengths used in the simulations are much bigger than in the sections shown here; far away in the tail region both shapes have concentrations $c=\Delta=0.6$. The growth velocities are $v/v_0=1.14$ (top) and $v/v_0=0.48$ (bottom).}
\label{fig4}
\end{center}
\end{figure}

A straightforward analysis of the stress state far behind the tip shows that among the possible configurations of new phases the energetically most favorable scenario are bicrystals, ${\theta=\pm 2\pi/3}$, as presented in Fig.~\ref{fig4}.
In the asymptotic tail region ${u_{xy}^{(1)}=0}$ and ${u_{xy}^{(2)}=\mp \varepsilon\sqrt{3}/2}$, where different signs correspond to different domains of the bicrystals.
The distribution of the strain component $u_{yy}$ and the fraction of the new phase $c$ can be readily found in the same way as before.
For example, Eq.~(\ref{c}) should be replaced by
\begin{equation}\label{ch}
c=\Delta=\frac{1}{1-2\nu}\left[\frac{4(1-\nu^2)}{E\varepsilon^2}(F^0_1-F^0_2)-\frac{1}{2}\right].
\end{equation}
The presence of the twin boundary with interfacial energy $\gamma_b$ requires also a modification of the characteristic length and velocity scales,
\[
L^*=8\frac{(1-\nu^2)(2 \gamma+\gamma_b)}{(1-2\nu)E\varepsilon^2}, \qquad v_0=\kappa (2\gamma+\gamma_b)/L^*
\]
compared to Eqs.~(\ref{c}) and (\ref{v0});
moreover, it leads to the existence of a triple junction in the tip region.

\begin{figure}
\begin{center}
\epsfig{file=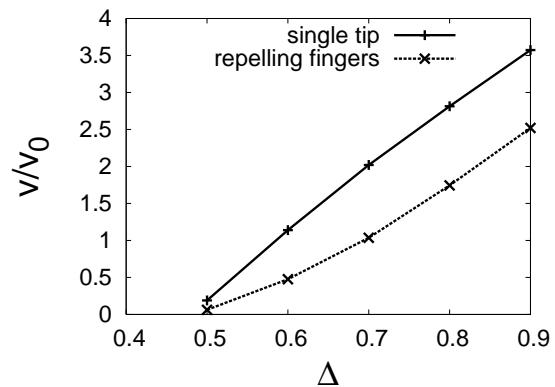, width=7.5cm}
\caption{Growth velocity of the different bicrystals for $L/L^*=5$. A slight discrepancy from the equilibrium point $\Delta_L$ is due to the accumulation of elastic energy in the transition region of the phase field, which leads to a renormalization of the surface energy.
We checked numerically that this effect is suppressed in the sharp interface limit, as expected.
}
\label{fig5}
\end{center}
\end{figure}

For a numerical phase field study of these twin structures we immediately take into account the symmetry of the appearing patterns and describe only either the upper or lower half of the strip.
At the symmetry plane the boundary conditions are $u_y=0$, $\sigma_{xy}=0$, and, for the specific case $\gamma_b\ll\gamma$ considered here, $\partial\phi/\partial y=0$.
This avoids more complicated multiphase descriptions which are in principle capable to describe the three different phases.
Although both patterns in Fig.~\ref{fig4} are energetically equivalent far away from the tip, symmetry is broken by the choice of the propagation direction.
For growth to the right, the orientation $\theta=2\pi/3$ in the upper and $\theta=-2\pi/3$ in the lower half, leading to propagation with a single tip (upper panel of Fig.~\ref{fig4}), is superior to the opposite case with repelling fingers (lower panel), as the growth velocity is higher, see Fig.~\ref{fig5}.

In summary, a simple sharp-interface kinetic model of strain influenced phase transitions has been developed together with its phase-field realization.
We obtained quantitative results for the steady-state growth of an elastic finger in a strip geometry and discussed the peculiar behavior of different pattern formation processes in this system.
Influence of additional conserved fields, e.g.~compositional and temperature field, can be subject of future investigations.

This work has been supported in part by the Deutsche Forschungsgemeinschaft under Grant SSP 1120 and by the CRDF Grant RUE1-1625-MO-06.
V.I.M. thanks Forschungszentrum J\"ulich for hospitality.

\end{document}